\documentclass[preprint2]{aastex}

\shorttitle{Lossless Astronomical Image Compression}
\shortauthors{Pence, Seaman, \& White.}

\begin{document}


\title{Lossless Astronomical Image Compression
and the Effects of Noise}


\author{W. D. Pence}
\affil{NASA Goddard Space Flight Center,
    Greenbelt, MD 20771}
\email{William.Pence@nasa.gov}

\author{R. Seaman}
\affil{National Optical Astronomy Observatories, Tucson, AZ 85719}

\and

\author{R. L. White}
\affil{Space Telescope Science Institute, Baltimore, MD 21218}



\begin{abstract}
We compare a variety of lossless image compression methods on a
large sample of astronomical images and show how  the compression
ratios and speeds of the algorithms are affected by the amount of
noise (that is, entropy) in the images. In the ideal case where the
image pixel values have a random Gaussian  distribution, the
equivalent number of  uncompressible noise bits per pixel is given
by $ N_{bits}=\log_2(\sigma \sqrt{12})$ and the lossless
compression ratio is given by   $R = {\sf BITPIX} / (N_{bits} + K)$
where {\sf BITPIX} is the bit length of the pixel values (typically
16 or 32), and $K$ is a measure of the efficiency of the
compression algorithm. We show that real astronomical CCD images
also closely follow these same  relations, by using a robust
algorithm for measuring the equivalent number of noise bits from
the dispersion of the pixel values in background regions of the
image.

We perform image compression tests on a large sample of  16-bit
integer astronomical CCD images using the GZIP compression program 
and using a newer FITS tiled-image compression method that
currently supports 4 compression algorithms: Rice, Hcompress, PLIO,
and the same Lempel-Ziv algorithm that is used by GZIP.  Overall,
the Rice compression algorithm strikes the best balance of 
compression and computational efficiency; it is 2--3 times 
faster and produces about 1.4 times greater compression than GZIP  
(the uncompression speeds are about the same).   The Rice algorithm
has a measured $K$ value of 1.2 bits per pixel, and thus
produces 75\%--90\% (depending on the amount of noise in the image)
as much compression as an ideal algorithm with $K$ = 0.
Hcompress produces slightly better compression  but at the expense
of 3 times more CPU time than Rice.   Compression tests on  a
sample of 32-bit integer images show similar results,  but the
relative speed and compression ratio advantage of Rice over GZIP is
even greater.  We also briefly discuss a technique for
compressing floating point images that converts the pixel values to
scaled integers.

The image compression and uncompression utility programs used in
this  study (called fpack and funpack) are publicly available from
the HEASARC web site.  A simple command-line interface may be used
to compress or uncompress any FITS image file.
\end{abstract}





\section{Introduction}

The size of astronomical data archives continues to increase
enormously, so it is in the interests of both data providers and
users to make use of the most effective image compression
techniques.  Compression reduces the storage media costs and the
network bandwidth needed to transmit the files to users. Image
compression also reduces the number of bytes of data that are
transfered to or from local disks during data analysis
operations.

The extensive literature on astronomical image compression can be
divided into 2 main categories:  lossy compression techniques, in
which some of the hard to compress information (ideally only
noise) is discarded, and lossless compression techniques where all
the information is preserved so that the original data can  be
exactly reconstructed from the compressed data.  Lossy compression
techniques \citep[e.g., ][ and references therein]{louys1999} can
provide higher compression than lossless techniques however
users must be careful to ensure that the required amount of
photometric and  astrometric precision in the compressed image is
preserved.   Lossless compression \citep[see ][ for a comprehensive
summary of previous work]{grunler2006} by definition preserves
all the information in the images and is often preferred or
required in situations where the data provider must be certain
that no information is lost, even if that means having to deal
with larger volumes of data.

In this paper we compare several lossless compression techniques
on a large sample of astronomical images. We describe the
different compression methods used in this study in section
\ref{s:tiledimage}.   Then in section \ref{s:noisecalc} we review
how the noise content of an image sets an upper limit on the
lossless compression ratio.   We use 2 sets of synthetic images
with known noise properties  to compare how well the  compression
ratios agree with the theoretical expectations.  This is followed
in sections \ref{s:shortimages} and \ref{s:intimages} with a
detailed comparison of how well the different algorithms perform
on actual 16-bit and 32-bit integer astronomical images taken mainly
with CCD detectors. Section \ref{s:floatimages} briefly
discusses compression methods for floating point images, and section
\ref{s:tilesize} discusses how the choice of tiling pattern
affects the compression performance.  Section \ref{s:summary} then
summarizes the main results of this study.  Finally, the appendix
gives a derivation of the formula for the equivalent number of
noise  bits in an image and shows how this relates to the seminal
work by \cite{shannon1948} on entropy in communication theory.

\section{Compression Methods}

\label{s:tiledimage} In this study we use a relatively new
compressed image format that is based on the FITS tiled-image
compression convention \citep{pence2000, seaman2007}.   Under this
convention, the image is first  divided into a rectangular grid  of
``tiles''.  Usually the image is tiled on a row by row basis, but
any other rectangular tile size may be specified.  Each tile of
pixels is then compressed using one of several available compression
algorithms (described below), and the compressed stream of bytes is
stored in a variable length array column in a FITS binary table. 
Each row of the FITS binary table corresponds to one tile in the
image.   Our software uses the CFITSIO library \citep{pence1999} to 
transparently read and write these compressed files as if they were
ordinary FITS images, even though they are physically stored in a
table format. One of the advantages of using this tiled image
convention, compared to the other technique currently used by most
observatories and data archive centers of externally compressing the
entire FITS image with the GZIP utility, is that the compressed FITS
image is itself a valid FITS file and the image header keywords
remain uncompressed, which provides faster  read and write access. 
Another advantage is that  each image in a multi-extension FITS file
is compressed separately and can be read without having to
uncompress the entire FITS file. Similarly, when reading a small
section of the image, only the corresponding tiles need to be
uncompressed.

The current implementation of this convention in the CFITSIO
library supports 4 lossless compression algorithms: Rice,
Hcompress, PLIO, and GZIP.  In principle, any other compression
algorithm, e.g., JPEG 2000 or bzip2, could be added in the
future.  We note however that the JPEG 2000 lossless algorithm,
JPEG-LS, uses the Golomb-Rice coding which is similar to the Rice
algorithm, and the bzip2 algorithm typically only provides a few
percent more compression than GZIP, but requires much more CPU
processing time,  especially when uncompressing the image
\citep{yang2002}.  The main features of each of the algorithms 
used in this study are described below.

{\bf Rice}: The Rice algorithm \citep{rice1993, white1998} is  very simple
(additions, subtractions, and some bit masking and shifts), making
it computationally efficient.  In fact, it has been implemented in
hardware for use on spacecraft and in embedded systems, 
and has been considered for use in compressing images from future space
telescopes \citep{nieto1999}.
In its usual
implementation, it encodes the differences of consecutive pixels
using a variable number of bits.  Pixel differences near zero are
coded with few bits and large differences require more bits.  
The algorithm
adapts to the noise by determining the number of pure
noise bits to strip off the bottom of the difference and include
directly in the output bitstream (with no coding).  The best value
for this noise scale is computed independently for each block of 16
or 32 pixels. With such short blocks, the algorithm requires little
memory and adapts quickly to any variations in pixel statistics
across the image.

{\bf Hcompress}: The Hcompress algorithm was
written to compress the Space Telescope Science Institute digitized
sky survey images \citep{white1992}. It involves (1) a
wavelet transform called the H-transform (a Haar transform
generalized to two dimensions), followed by (2) an optional
quantization that discards noise in the image while retaining the
signal on all scales, followed by (3) a quadtree coding of the
quantized coefficient bitplanes. In this study we omitted the
quantization step, which makes Hcompress lossless. The H-transform 
computes sums and differences within pixel blocks, starting with
small 2x2 blocks and then increasing by factors of two to 4x4, 8x8,
etc., blocks. This is an exactly reversible, integer arithmetic
operation, so a losslessly encoded set of the H-transform
coefficients can be uncompressed and inversely transformed to
recover the original image. The H-transform  can be performed
in-place in memory and requires enough memory to hold the original
image (or image tile).  To avoid overflow problems when summing  the
pixel values, the memory array is expanded by a factor of 2 so that
each pixel has twice as many bits as in the original image. The
Hcompress bitplane coding, which proceeds by first compressing the
most significant bit of each coefficient (mostly zeros) and working
down to the least significant bit (usually noise), has the
effect of ordering the image description so that the data stream
gives a progressively better approximation to the original image as
more bits are received.  This was used to create an efficient
adaptive scheme for image transmission \citep{percival1996}.

{\bf PLIO}: The IRAF \citep{tody1993} Pixel List I/O (PLIO)
algorithm was developed to store integer image masks in a compressed
form.   This special-purpose run-length encoding algorithm is very
effective on typical masks consisting of isolated high or low values
embedded in  extended regions that have a constant pixel value.  Our
implementation of this algorithm only supports pixel  values in the
range 0 to $2^{23}$. Because of the specialized nature of the PLIO
algorithm, we only  discuss its use with compressing data masks, in
section \ref{s:datamask}.

{\bf GZIP}: \label{s:GZIP} The popular GZIP file compression
utility \citep{gailly} is the defacto standard compression method
currently used in the astronomical community.  Nearly all major
observatories and data archive centers distribute their data as
GZIP compressed files.  For this reason, GZIP serves as the baseline
of comparison for the other compression methods in our study.
GZIP uses a variation of the Lempel-Ziv algorithm \citep{ziv1977}
to build a dictionary of repeated
sequences of bytes occurring in the input and using a short code for
each sequence.  The most important distinguishing characteristic of
GZIP compared to the other compression algorithms used in this study
is that GZIP treats each 8-bit byte of the input data stream as an
independent datum, whereas the other compression methods operate on
the numerical value of the input image pixels as multi-byte
quantities. This puts GZIP at a distinct disadvantage when
compressing astronomical images with 16-bit or 32-bit pixel values
because, unlike the Rice and Hcompress algorithms, GZIP cannot use
the numerical difference between adjacent pixels as a means of
improving the compression. As a result, it becomes less effective when
increasing noise makes repeated bit patterns less common. 

It should be noted that the GZIP algorithm has a user-selectable 
parameter for  fine tuning the  trade off between speed and
compression ratio, where a value of 1 gives the fastest compression
at the expense  of  file size and 9 gives the highest compression at
the expense of speed.  Using the fastest value of 1 instead of the
default value of 6  can increase  the compression speed by a factor
of 2 or more while only increasing the compressed file size by a few
percent,  therefore we have used this in all the speed comparison
tests in this study.  One small side effect,  however, is that it
increases the subsequent image uncompression time by about 10\%.

Within this study, the GZIP algorithm is used in 2 different
processing contexts which have significantly different speeds. In
the first context, the GZIP program on the host computer is used to
externally compress the FITS image, and in the other context the
GZIP algorithm is used within the FITS tiled image convention to
compress each image tile. The numerical algorithm is identical in
both cases, however  the host GZIP program only takes about half as
much CPU time as the tiled GZIP method  to compress the same image.
This difference is mainly due to the fact that the host GZIP program
can more efficiently read and write the input and output files as
sequential streams of bytes, whereas the tiled image compression
method  requires random access to the FITS files, which in turn
requires that the input and output data be copied to intermediate
storage buffers in memory. As will be demonstrated later, in spite
of this extra processing overhead  the tiled image Rice algorithm
can still compress  images several times faster than the host GZIP
program. 

\section{The Effect of Noise on Lossless Image Compression}
\label{s:noisecalc}

The fundamental principle that limits the amount of lossless image
compression is the well known fact that noise (that is,
entropy) is inherently incompressible. In this section we
quantitatively  demonstrate how the amount of noise in an image can 
be measured and used to predict the lossless image compression ratio.

In order to study the affect of noise on image compression in a
large sample of images, we need an algorithm that can reliably
estimate the amount of noise in any given image without manual
interaction or iterations.  Such an algorithm was recently
developed  by \cite{stoehr2007} for measuring the signal-to-noise
in  spectroscopic data.  It is based on the median absolute
deviation (MAD) method of quantifying the variations in a sample of
values, which is less sensitive to the presence of outlying values
than other simple statistical measures such as the standard
deviation. To help mitigate against the possible presence of 
correlations in the noise between adjacent pixels in an image,
\cite{stoehr2007} developed a series of higher-order MAD  equations
that use the differences between the values of  every other pixel in
each row of the image. In our study we have  adopted their 3rd order
MAD equation (as they did as well) as a good compromise between
simplicity and accuracy.  This MAD value for each row of the
image is given by:
\begin{equation}
\sigma = 0.6052 \times {\sf median} ( -x_{i-2} + 2 x_i - x_{i+2} ) 
\label{eq:noise3}
\end{equation}
where $i$ is the vector index of the pixel within each row of the
image, and $x$ is the value of the indicated pixel. 
The median value is computed over all the pixels in the
row. (A C-language implementation of this algorithm is
available in our image compression software which can be downloaded
as described at the end of this paper).  The use of the median in
this formula is very effective at eliminating any outlying  large
deviations, for instance, close to the  images of star and galaxies
in typical astronomical images. Also, this formula is insensitive to
large-scale gradients in the mean background across the image.  As a
result this formula provides a good measure of the pixel variations
relative to the {\em local  background} regions of the image,  in
between any localized brighter objects in the image.

It is easier to understand how noise affects image compression by
considering  a hypothetical image that has {\sf BITPIX} bits per
pixel (where {\sf BITPIX} is usually 8, 16, or 32 in astronomical
images) and where the lowest $N$  bits of each pixel contain
only noise (i.e., where each bit is  randomly assigned a value of 0
or 1), and the higher order  ${\sf BITPIX} - N$ bits are all
set to 0.  A histogram of the integer pixel values in such an image
will have a  flat distribution ranging from 0 to $2^N -1$.   Since
the $N$ noise bits in each pixel are incompressible, the maximum
possible compression ratio that can be achieved, if all the
remaining bits are infinitely compressed to 0, is given simply by
\begin{equation}
R = {\sf orig\_size} / {\sf comp\_size} = {\sf BITPIX} / N_{bits}
\end{equation} 
where $N_{bits}$ is the (average) number of noise bits per
pixel in the image.  $N_{bits}$ is equivalent to the
\cite{shannon1948} entropy expressed  in bits (see Appendix). In
practice, no actual algorithm can infinitely compress all the
non-noise bits, and instead can only compress them, on average, down
to $K$ bits per pixel. This $K$ parameter can be viewed as a measure
of the efficiency of a compression algorithm, where better
algorithms have smaller values of $K$. Thus, the compression ratio
achieved by actual compression algorithms is given by
\begin{equation}
 R = {\sf BITPIX}/(N_{bits} + K)
\label{eq:ratio}
\end{equation}

Unlike in the above example, the noise in astronomical images is
usually not neatly confined to the lowest $N$ bits of each
pixel. The actual noise distribution in the types of astronomical 
CCD images studied in this paper is a complex sum of the Poisson
photon noise plus multiple noise sources in the detector, such as
from the charge transfer  and  read-out electronics.  In practice,
the sum of all these noise contributions near to the background
level in a CCD image can be closely approximated by a Gaussian. 
(Fitting a Gaussian to the histogram of the pixel values is a common
image processing  technique for measuring the mean background
level.) If one assumes that the noise in the image pixels has a
Gaussian distribution, with standard deviation $\sigma$, then we
show in the appendix that the {\em equivalent} number of noise bits
per pixel in the image is given by
\begin{equation}
N_{bits} = {\log}_2 (\sigma \sqrt{12}) = \log_2 (\sigma) + 1.792 
\label{eq:nbits}
\end{equation}
and the expected compression ratio for this image is then given by
\begin{equation}
R = {\sf BITPIX}/(\log_2 (\sigma) + 1.792 + K)
\label{eq:ratio2}
\end{equation}
Thus, in summary, we can use the MAD equation~\ref{eq:noise3} to
estimate the noise in the background regions of an image, and then
use  equation~\ref{eq:ratio2} to predict how much that image can be
losslessly compressed.  For example, a 16-bit integer image with
MAD  $\sigma = 18$ contains  about 6 bits per pixel of
uncompressible noise and will compress by a factor of 2.3 when using
a compression algorithm that has $K$ = 1.

\subsection{Compression of Synthetic images with Known Noise Properties}
\label{s:synthImages}

Before measuring the performance of the different compression 
algorithms on real astronomical images, it is instructive to first
study the behavior of the algorithms on 2 sets of synthetic 16-bit
and 32-bit integer images that by construction have noise properties
that match the 2 cases discussed in the previous section.  In the
first set of synthetic images the lowest $N$ bits of each
integer pixel were randomly assigned a value of 0 or 1 and the upper
{\sf BITPIX} -- $N$ bits are all set to 0. In the second set of
synthetic images, the pixels were assigned values randomly selected
from a Gaussian distribution, with  $\sigma$ ranging from 1.0 to
500.  (We also added a constant offset to the pixels in these
images  to avoid negative values, however  we verified, as expected,
that the  magnitude of this offset has no effect on the performance
of the compression algorithms.)  The equivalent number of noise bits
per pixel in this second set of images is given by
equation~\ref{eq:nbits}.

\begin{figure}
\plotone{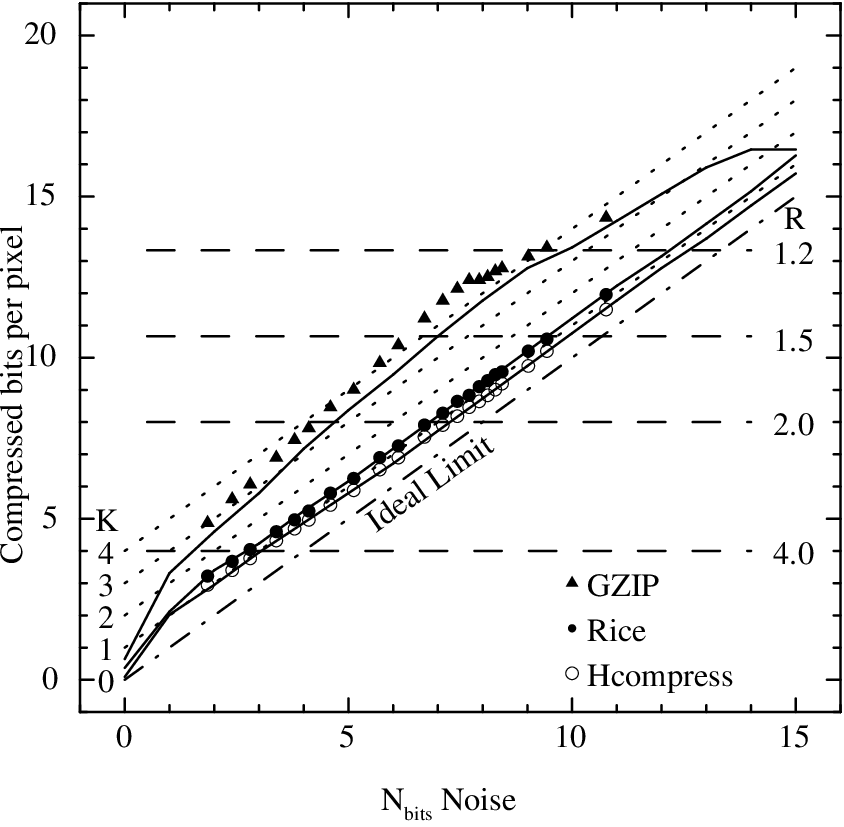}

\caption{Plot of the compressed bits per pixel versus the number 
of noise bits in 16-bit synthetic images.  The solid lines
represent the images that have $N_{bits}$ of uniformly
distributed noise, and the symbols represent the images that have
Gaussian distributed noise. The diagonal dotted lines have
constant  $K$ values; the lowest line with $K$ = 0 is the
theoretical limit. The horizontal dashed lines show the image
compression ratio corresponding to the scale on the Y axis.
\label{f:compbit_v_noise16}
}
\end{figure}

\begin{figure}

\plotone{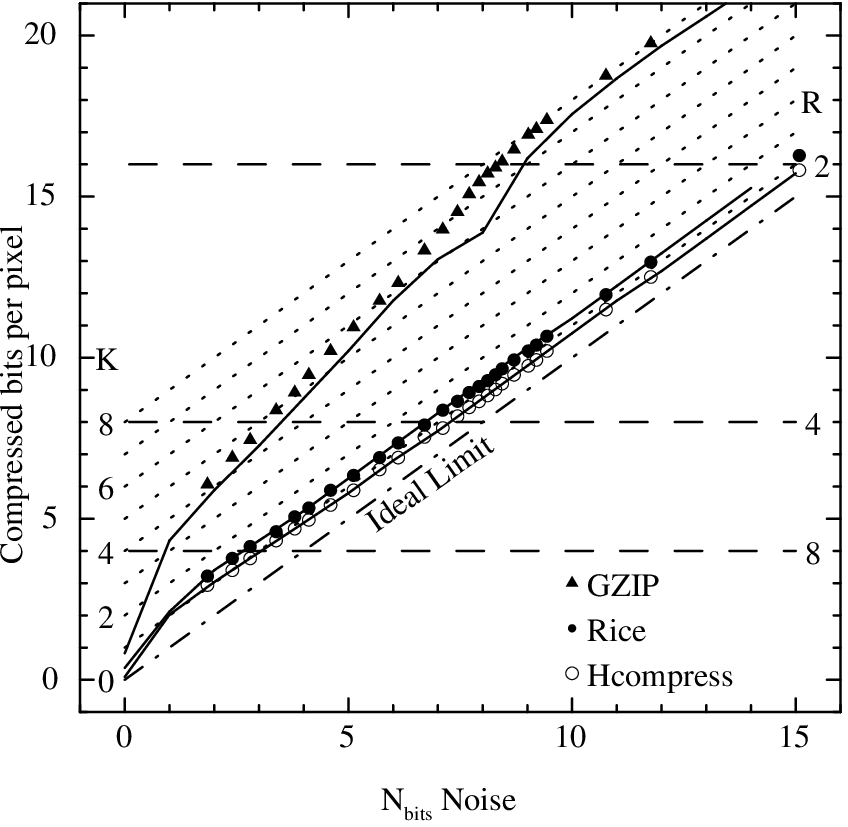}

\caption{Same as Figure 1, except for the 32-bit integer synthetic noise images.
\label{f:compbit_v_noise32}
}
\end{figure}

We then measured how the compression ratio of these synthetic images
varies as a function of the number of noise bits (or equivalent
noise bits) in the image when using the 3 general-purpose
tiled-image compression algorithms, Rice, GZIP, and Hcompress. 
Instead of directly plotting the compression ratio, $R$, it is
more informative to plot the reciprocal quantity ${\sf BITPIX}/R$ as a function of  the number of noise bits per pixel, because in
this coordinate frame the lines of constant $K$ value have a
slope = 1.0 and a Y-intercept = $K$.   The results for the
16-bit integer synthetic images are shown in Figure
\ref{f:compbit_v_noise16} where the solid lines are derived from the
images with uniformly distributed $N$ bits of noise, and the
circular or triangular points are derived from the other set of
images with a Gaussian noise distribution.  The lowest line and set
of points is derived using Hcompress, the middle ones using Rice,
and the upper ones using GZIP.  The diagonal dotted lines represent
the different constant $K$ values, and the horizontal dashed
lines show the image compression ratio that corresponds to the
compressed bits per pixel scale on the Y axis.

In the case of the Hcompress and Rice algorithms, it can be seen
that the points lie almost exactly on top of the corresponding line.
This provides strong empirical confirmation of the $\sqrt{12}$ 
horizontal offset (derived in the appendix) that is needed to
produce this good agreement.  A least squares solution for the
offset that produces the best agreement between the 2 sets of images
gives $1.80 \pm 0.02$ for Hcompress, and $1.76 \pm 0.02$ for Rice,
in excellent agreement with the expected value of $\sqrt{12} =
1.792$. It can also be seen that the Hcompress and Rice lines have a
slope very close to 1.0,  which means that the $K$ compression
efficiency factor for these 2 algorithms is nearly constant and is
independent of the amount of  noise in the image.  The best fitting
$K$ values are $0.78 \pm 0.02$ bits per pixel for Hcompress and
$1.18 \pm 0.02$ for Rice. Close inspection of the figure shows
however that the $K$ value does increase slightly for images
with less than about 5 bits of noise.  We attribute this trend to
the  fixed-size disk space ``overhead'' in the FITS tiled-image
compression format which becomes relatively more significant as the
image becomes more highly compressed. (See also the discussion  in
the appendix of another non-linear effect at small values of $N$).

It is apparent that GZIP behaves very differently from Rice or
Hcompress in Figure \ref{f:compbit_v_noise16}.  GZIP cannot be
parameterized with a single value of $K$, and instead ranges
from about 2 to 5 bits per pixel, depending on the amount and
distribution of the noise in the image.   Unlike the other 2
algorithms,  GZIP does not compress the 2 types of synthetic noise
images equally well; it  compresses the images in which the noise is
confined to the lowest $N$ bits better than the images with a
Gaussian noise distribution.   It is interesting that $K$
appears to reach a maximum at $N_{bits} = 8$  which is where
the noise propagates into the more significant byte of the 2-byte
pixel values.  We attribute most of these differences between GZIP
and the other 2 algorithms to the fact that Hcompress and Rice treat
each 16-bit pixel value as a single integer number, whereas GZIP
treats each 8-bit byte as an independent datum.

The equivalent plot for the synthetic 32-bit integer images is shown
in  Figure \ref{f:compbit_v_noise32}.  The relations for Rice and
Hcompress are virtually identical to those for the 16-bit integers,
and in particular, the  $K$ values are the same.   As was the
case with 16-bit images, GZIP behaves quite differently and has a
variable $K$ value that approaches 8 bits per pixel
for the noisiest images (i.e.,  has a full byte per pixel of
overhead compared to the ideal $K$ = 0 algorithm).

\section{Compression of 16-bit Astronomical Images}

\label{s:shortimages}
In this section we examine how well the different compression
methods perform on real 16-bit integer astronomical images. The
primary data set used in these tests is the set of images that were
taken during the night of 27 -- 28 July 2006 at  Cerro Tololo
Inter-American Observatory using the Mosaic CCD camera.  This data
set contains a variety of different types of images that are
typically taken during an observing session,  including 0s exposure
bias frame images, heavily exposed flat-field images,  short
exposures (10s -- 30s) of bright calibration stars, and longer
exposures (500s -- 600s)  containing randomly distributed images of
stars, faint galaxies, and diffuse emission.  These are typical of
the types of images obtained by  many sky-survey projects. The
Mosaic camera contains 8 individual CCD detectors, and each detector
has 2 amplifiers that read out half of the chip each.   Thus, every
exposure with this camera results in a  FITS file containing 16
image extensions that are each 1112 by 4096 pixels in size. In
total, this data set consists of 102 FITS files containing 1632
separate FITS image extensions.  

To complement this large homogeneous set of images taken with the
Mosaic  camera, we also included in our test sample a smaller set of
16-bit integer images taken with a variety of other instruments. 
First we included the suite of test images that were collected by
\cite{murtagh1989} for use in testing  image processing techniques.
This sample of images has been used in many previous studies 
\citep{richmond1995, sabbey1999, grunler2006} and remains available
from ftp://iraf.noao.edu/iraf/extern/focas.std.tar.Z. In order to
further increase the diversity of images in our sample,  we included
9 other  more recent deep-sky images obtained from the archives of
the Hubble Space Telescope, ESO, and the Anglo-Australian
Observatory.

We compressed and uncompressed each of these images using the Rice,
GZIP, and Hcompress algorithms supported by our tiled-image
compression software,  and in each case recorded  the compression
ratio and the  elapsed compression and uncompression CPU times. 
These same parameters were measured using the GZIP utility  program
on the host computer to externally compress and uncompress the
images. These host GZIP tests were  performed on  a single FITS
image extension instead of on the whole multi-extension file,  to be
comparable with the tiled-image compression tests which also operate
on a single image extension at a time.  We also calculated the MAD
pixel dispersion in the background regions of each image from
Equation \ref{eq:noise3} and the corresponding equivalent number of
noise bits from Equation \ref{eq:nbits}.  

\subsection{Compression Ratio versus Noise}

\begin{figure}
\plotone{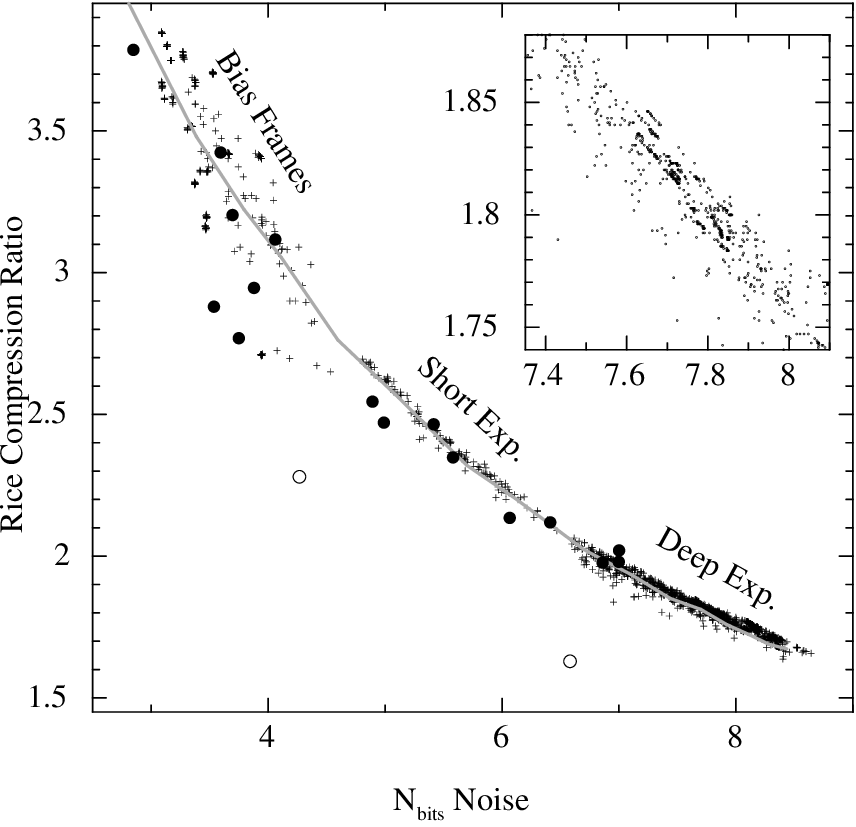}

\caption{The Rice compression ratio plotted as a function of the
equivalent number of noise bits in 16-bit integer astronomical images. 
The + symbols show the Mosaic CCD camera data set, and the larger symbols
show the comparison set of images taken with other instruments.  The gray
line is derived from both sets of synthetic noise images.   The insert is
a magnified view of the lower right section of the larger plot showing the
bands from the 16 different detectors. 
\label{f:rice_v_noise} }
\end{figure}

One of most striking results of this study, as  shown in Figure
\ref{f:rice_v_noise}, is the very tight correlation between the
compression ratio and the measured number of equivalent noise bits
in  the Mosaic camera images (plotted with small + symbols).   The
gray line going through these points is derived from the  2 sets of
synthetic noise images  discussed in the previous section and
corresponds closely to the $K$ = 1.2 line in Figure
\ref{f:compbit_v_noise16}. (The compression ratio $R$, is
plotted here, rather than the reciprocal quantity ${\sf BITPIX} /
R$ that was use in the previous figures, because $R$ is
the quantity of more direct interest to most users.) The larger
points in this figure are derived from the comparison sample of
images, and show that they also generally follow the same relation
as the Mosaic images, but with somewhat larger scatter.

The close agreement between the Mosaic camera images and the
synthetic noise images in Figure \ref{f:rice_v_noise} demonstrates
that the presence of ``objects'' in the astronomical images (e.g.,
the stars and galaxies) has little impact on the compressibility of
these images.   This is also demonstrated by the continuity between
the Mosaic camera bias and flat field images (which contain no
objects) and the images of the sky. This is simply a result of the
fact that most of the pixels in the Mosaic camera sky images have
values close to the background level and only a few percent of the
total image area is significantly affected by the brighter
objects.  Thus, even if the algorithms do not compress the regions
close to these bright objects very effectively,   the overall
compression ratio of the image will still mainly be determined by
the compressibility of the background regions.  Since our MAD noise
estimation algorithm (equation \ref{eq:noise3}) measures
the noise in the background regions, it is an excellent predictor
of the overall image compression ratio of these images.

Obviously there can be exceptions to this general rule if the
objects in the image occupy a significant fraction of the image
area.  This is the case for the 2 images in our comparison sample
that are plotted with open circles in  Figure
\ref{f:rice_v_noise}.  These are a pair of long and short exposures
of the NGC 3201 globular cluster \citep[the ngc0001.fits and ngc0002.fits
files from the ][ image suite]{murtagh1989}   which contain an
unusually dense pattern of star images. In cases like this the
overall image compression ratio can be significantly less than what
would be predicted simply from the pixel variations in the local
background regions between the stars images.  

It is also interesting to note in Figure \ref{f:rice_v_noise} that
the Mosaic camera images are segregated into 3 distinct groups that
correspond to the bias frame images (containing the least amount of
noise), the short exposures of calibration stars (the middle
group), and the deeper exposures of the sky and flat fields
(containing the most noise).  This grouping is just a reflection of
the fact that these three different types of images have distinctly
different  average pixel values, and the noise (mainly from Poisson
statistics) scales roughly as the square root of that value. 

The insert in Figure \ref{f:rice_v_noise} shows a magnified section
of the data in which it can be seen that the points tend to lie
along distinct bands which correspond to the 16 different detectors
in the Mosaic camera. The points from a single detector are tightly
correlated and thus most of the scatter seen in this figure is due
to the systematic differences between image detectors.  We will
come back to this effect in the discussion of the NEWFIRM images 
in the following section which show an even more pronounced banding
pattern.

\subsection{Comparison of Different Compression Algorithms}
\label{s:compare16bit}
\begin{figure}
\plotone{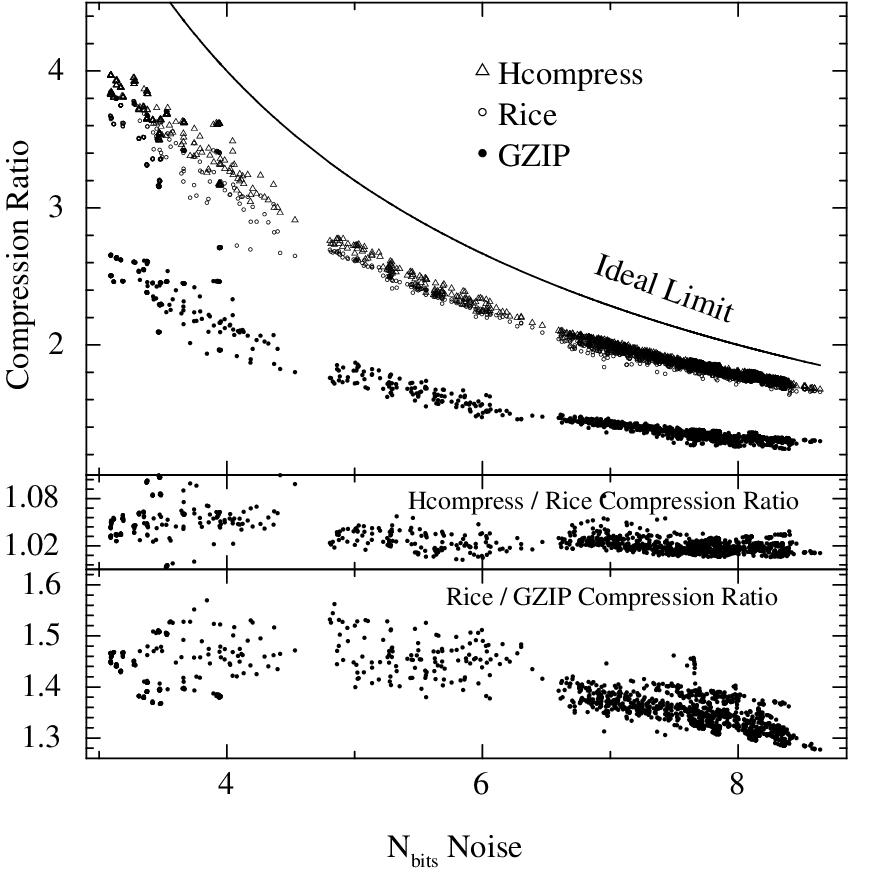}

\caption{The compression ratios for the different algorithms as a
function of the equivalent number of noise bits in the 16-bit
images (upper panel).  The solid line shows the upper limit for an
ideal compression algorithm with $K$ = 0.  The middle and lower
panels compare the Rice compression ratio to that of Hcompress and
GZIP, respectively.
\label{f:all_v_noise}
}
\end{figure}

Figure \ref{f:all_v_noise} compares the compression ratios achieved
by all three algorithms plotted as a function of the equivalent
number of noise bits in the Mosaic camera CCD images. As can be
seen, Rice and Hcompress achieve very similar compression ratios
that are about 1.4 times greater than when using Gzip. The bottom
panel shows that this ratio varies from about 1.5 for images with
low to moderate amounts of noise, to about 1.3 for the noisest
images. The middle panel shows that Hcompress produces about 2\% to
5\% better compression than Rice, but as discussed below, this
small gain is usually not worth the much higher required CPU times.

The upper solid curve in Figure \ref{f:all_v_noise}  shows the
theoretical maximum compression ratio, given by  {\sf BITPIX} /
$N_{bits}$, of an ideal lossless algorithm that infinitely
compresses all the non-noise bits in the image  (i.e., an algorithm
with $K = 0$).  The Rice and Hcompress algorithms have measured
$K$ values of 1.2 and 0.8  bits per pixel,  respectively, and
thus produce about 75\% to 90\% of the ideal amount of compression,
depending on the noise level.  

\begin{figure}
\plotone{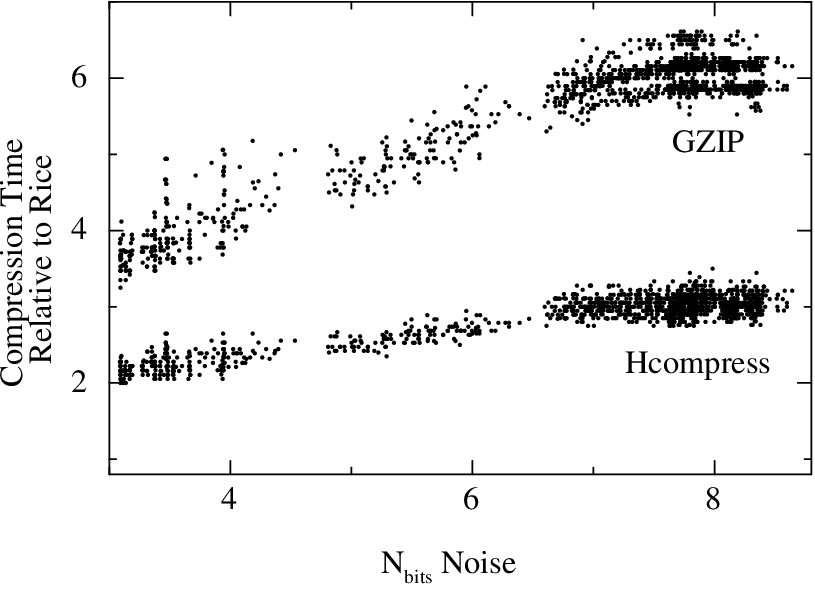}

\caption{CPU time needed to compress 16-bit integer  FITS images
using  GZIP (top) or Hcompress (bottom) relative to the time when
using the Rice algorithm. The horizontal banding of the points  is due
to the finite time resolution of the CPU measurements.
\label{f:mosaic_relative_pk_times}
 }
\end{figure}

\begin{figure}
\plotone{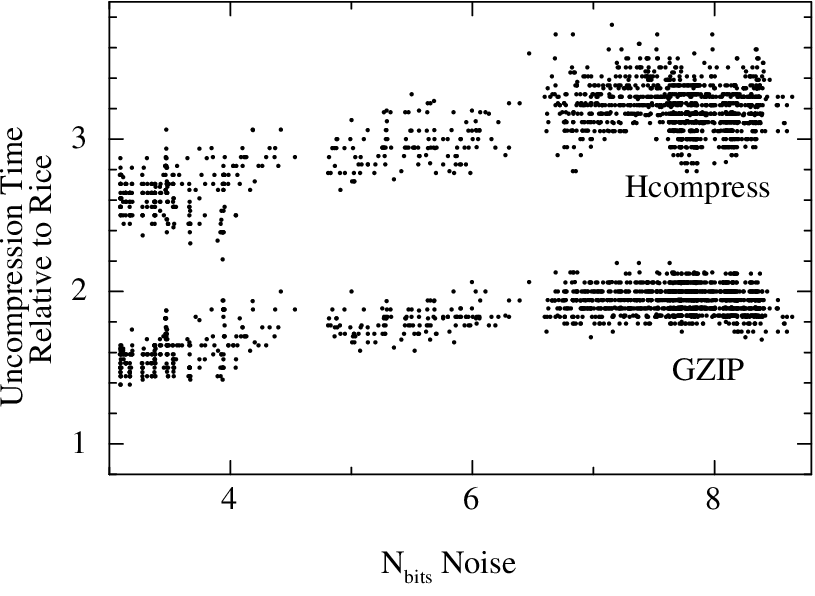}

\caption{CPU time needed to uncompress 16-bit integer  FITS images
using GZIP (top) or Hcompress (bottom) relative to the time when using
the Rice algorithm.  The horizontal banding of the points  is due to
the finite time resolution of the CPU measurements.
\label{f:mosaic_relative_unpk_times}
 }
\end{figure}

The relative compression and uncompression speeds\footnote{ The
timing measurements in this article are based on CPU time and not on
the total elapsed processing time, which includes the time needed
to access data on magnetic disk. The latter time should be larger,
however, it is difficult to measure consistently because of the
sophisticated data caching techniques used by modern computer
systems.}
of the different methods are shown in Figures
\ref{f:mosaic_relative_pk_times}  and 
\ref{f:mosaic_relative_unpk_times}.  
The Rice compression algorithm is 2--3 times faster than Hcompress 
(depending on the amount of noise) and 4--6 times faster than
tiled-GZIP (or about 2--3 times faster than the host GZIP utility,
as explained in section \ref{s:tiledimage}).
When uncompressing images, Rice is 2.5--3 times faster than
Hcompress, and 1.6 to 2 times faster than tiled-GZIP (or about the 
same speed as the host GUNZIP utility).

The mean compression ratios and the mean compression and uncompression  CPU
times for all 1632 Mosaic camera images are given in Table 1 for each of
the compression methods. The CPU times are relative to the Rice algorithm. 
As a benchmark reference,  a Linux PC with a 2.4 GHz AMD Opteron 250 dual
core processor (using only one of the processors) can  Rice-compress  a 50
MB 16-bit integer FITS image in 1 second.  Uncompressing this image  also
takes about 1 second.

\begin{deluxetable}{lcccc}
\tablecaption{Compression statistics for 16-bit integer images
\label{t:16int}
}
\tablewidth{0pt}
\tablehead{
\colhead{ } & \colhead{Rice} & \colhead{Hcompress} & \colhead{Tiled-GZIP} & 
\colhead{Host-GZIP} 
}
\startdata
Compression Ratio      & 2.11 & 2.18 & 1.53  & 1.64 \\
Compression CPU time   & 1.0  & 2.8  & 5.6   & 2.6 \\
Uncompression CPU time & 1.0  & 3.1  & 1.9   & 0.85 \\
\enddata
\end{deluxetable}

\subsection{Special case: Data Masks}
\label{s:datamask}

Data mask images are often used in processing environments as a means of
flagging special conditions that affect the corresponding pixels in an
associated astronomical image.   The pixel values in a data mask contain
essentially no noise,  so the general relationship between noise and
compression ratio as discussed  previously does not apply.   Instead of being
limited by noise, the maximum compression ratio of a data mask is  determined
by internal limits within each compression method.   With the Rice algorithm,
for example, each block of 32 image pixels (= 64 bytes) can at most  be
compressed to  a single 4-bit code value for a maximum compression ratio of
128.  Similarly, the maximum compression  ratio is about 200 for GZIP and
about 700 for both Hcompress and PLIO. Other factors, such as the image tile
size, can  further limit the compression ratio, but in practice, large
compression ratios of 50 or more can often be achieved.

In many cases, choosing the algorithm that produces the very highest
compression ratio is of secondary importance  because the masks compress so
well with any algorithm that the size is insignificant compared to the rest
of the associated data set. The compression and uncompression speed of the
algorithm can be a more significant consideration in this case.  Our tests
on  a sample of data masks show that Rice and the IRAF PLIO algorithm
are the fastest, but GZIP and Hcompress are less  than a factor of 2 slower. 
Overall, PLIO provides the best combination of compression ratio and speed
when compressing data masks, but the difference with the other algorithms is
not great.  In practice it may be simplest to just use the same compression
algorithm on the data mask as is used on the associated astronomical image.

\section{Compression of 32-bit Integer Images}
\label{s:intimages} 

Astronomical images in 32-bit integer format are less frequent
than 16-bit integer images because few imaging detectors produce more
than  16 bits per pixel of precision.  One of the few such instruments in
general use is the NEWFIRM near-infrared camera at the Kitt Peak National
Observatory. In order to measure the performance of the different
compression methods on 32-bit integer images, we used a sample of FITS
images taken with this camera during the night of 24 -- 25 February 2008.
These images are similar to the Mosaic data set, and include  bias frames,
flat fields, short exposures of calibration stars, and longer  exposures of
the sky containing images of stars and faint galaxies. The NEWFIRM
instrument contains a mosaic of 4 imaging detectors, each of which is 2112
by 2048 pixels in size.  There are 447 NEWFIRM observations in our data
sample, giving a total of 1788 separate images.  

\begin{figure}
\plotone{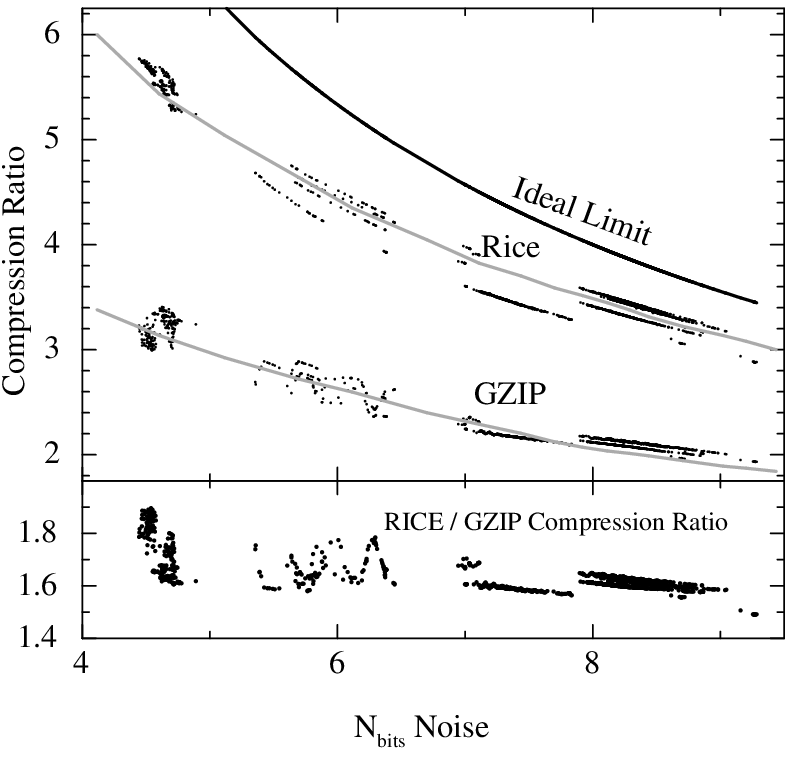}

\caption{ The Rice and GZIP compression ratios as a function of the
equivalent number of noise bits in the 32-bit images (top panel). 
The upper solid line shows the limit for an ideal compression
algorithm with $K$ = 0.  The gray lines going though these points
are derived from synthetic images with Gaussian distributed noise. The
lower panel compares the Rice compression ratio to GZIP.
\label{f:newfirm_ratio_v_noise}
}
\end{figure}

\subsection{Comparison of Different Compression Algorithms}
\label{s:compare32bit}

We repeated the same analysis as was done in the previous section on the
16-bit images to measure the compression ratios and the CPU times required
to compress and uncompress each of the 32-bit integer NEWFIRM images with
each different compression method. Figure
\ref{f:newfirm_ratio_v_noise} shows how the Rice and tiled-GZIP compression
ratios  depend on the measured number of equivalent noise bits in each
image.   (The points for the Hcompress algorithm have been omitted for
clarity because  they lie only slightly above the Rice points.)  This figure
is similar to the corresponding Figure \ref{f:all_v_noise} for 16-bit 
integer images, except that the compression ratios are twice as large, given
the same amount of noise.  This is to be expected from equation
\ref{eq:ratio} and is a natural consequence   of the fact that a 32-bit
integer image is a factor of 2 larger than a similar 16-bit image, but if
they both have the same equivalent number of noise bits per pixel then the
compressed images will be identical in size when using algorithms like Rice
and Hcompress that have a constant $K$ value.

This 2:1 relationship in compression ratios does not hold for the GZIP 
algorithm because $K$ is larger for 32-bit integer images, as can be
seen by comparing Figures \ref{f:compbit_v_noise16} and
\ref{f:compbit_v_noise32}.  Thus, the compression ratio of a 32-bit image
when using GZIP is only about 1.6 times greater then that of a 16-bit image
with the same equivalent number of noise bits. 

As was also the case for the 16-bit astronomical images, the different 
types of images are segregated into separate regions of Figure
\ref{f:newfirm_ratio_v_noise} because the mean pixel value, and hence the
amount of noise, is distinctly different. The clump of points with 
$N_{bits} < 5$ are the 0s exposure `bias' calibration exposures,  the points
with $5 < N_{bits} < 7 $  are the short exposures of calibration
stars, and  the remaining points with larger noise values are the deep sky
images and the heavily exposed flat field images.

The upper solid curve in Figure \ref{f:newfirm_ratio_v_noise} shows the 
maximum possible lossless compression ratio that would be achieved by an 
ideal compression algorithm that has $K = 0$.   Hcompress and Rice, with  
$K$ values of 0.8 and 1.2, respectively, are within 75\% to 90\% of
this theoretical limit, depending on the amount of noise, just as with
16-bit integer images.

Finally, one other prominent feature in Figure \ref{f:newfirm_ratio_v_noise}
is that the points are  split into  distinct bands that correspond to the 4
different detectors in the NEWFIRM camera.  This is similar to the banding
seen in the 16-bit Mosaic camera image, but on a larger scale. Unlike the
CCD detectors in the Mosaic camera, which are closely matched in  image
quality, the 4 infrared imaging devices in the  NEWFIRM camera have
distinctive characteristics.   One notable feature is that these images show
faint streaks in the background, and the streaks run vertically in 2 of the
chips and horizontally in the other 2 (as a result of the way the chips are
oriented in the camera).   Since our MAD algorithm calculates the noise on a
row by row basis, the noise value is larger in the  cases where the
rows cut across the grain of the streak pattern. This causes a systematic
displacement of the points from the different chips in the figure.  

Naively, one would expect the displacements would fall along the gray line
in the figure that is derived from synthetic images  with uniformly or
Gaussian distributed noise, so that  any change in the noise level causes a
corresponding change in the compression ratio according to
equation~\ref{eq:ratio2}.  But as can be seen, there are instances where 2
images taken with different detectors have the same equivalent number of
noise bits but have systematically different compression ratios (i.e., there
is a vertical displacement between the points from the different chips in
the figure). This indicates that these images have properties that differ in
some way from the assumptions that went in to the derivation of
equation~\ref{eq:ratio2}. As a simple example, if one were to add a small
constant value to all pixels in the even-numbered columns of an image, this
would  increase the dispersion in the pixel values along each row of the
image, causing the compression ratio to decrease, however the calculated MAD
noise value would remain unchanged because that calculation is based on the
differences between every other pixel value in the row. More generally, any
deviations from the assumed pure Gaussian distribution in the image pixel
values can lead to systematic offsets in the calculated equivalent number of
noise bits, or in the efficiency of the compression algorithm (i.e., the
$K$ value), or both.  At some level every physical device will show
deviations from an ideal detector.  These effects are relatively large in
the NEWFIRM detectors,  but the banding seen in the insert in Figure
\ref{f:rice_v_noise} indicates that they are also present in the 16-bit
Mosaic camera image at a smaller level.  This effect probably also accounts
for  some of scatter seen in that figure in the comparison sample of images
taken with other instruments.  If one were to analyze a larger sample of
images from each of those instruments, they  would likely fall within a
fairly narrow region in that figure. 

\begin{figure}
\plotone{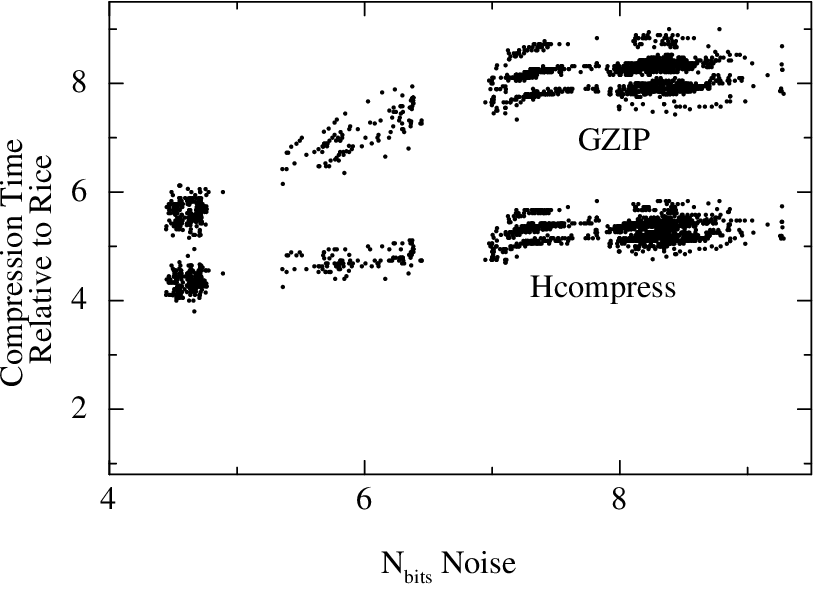}

\caption{
CPU time needed to compress 32-bit integer  FITS images using  GZIP
(top) or Hcompress (bottom) relative to the time when using the Rice
algorithm. The horizontal banding of the points  is due to the finite
time resolution of the CPU measurements.
\label{f:newfirm_relative_pk_times}
}
\end{figure}

\begin{figure}
\plotone{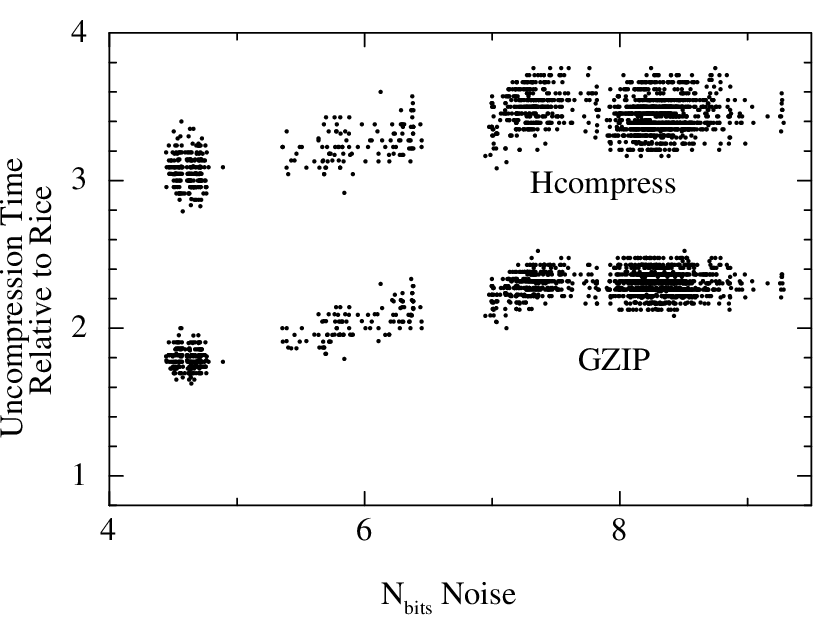}

\caption{
CPU time needed to uncompress 32-bit integer  FITS images using  GZIP
(top) or Hcompress (bottom) relative to the time when using the Rice
algorithm. The horizontal banding of the points  is due to the finite
time resolution of the CPU measurements.
\label{f:newfirm_relative_unpk_times}
 }
\end{figure}

A comparison of the CPU times required to compress  and uncompress the
32-bit integer images with GZIP or Hcompress, relative Rice, is shown  in
Figures \ref{f:newfirm_relative_pk_times} and
\ref{f:newfirm_relative_unpk_times}. The average compression ratios and the
relative compression and uncompression  CPU times for all 1788 NEWFIRM
images are also summarized in  Table \ref{t:32int}.   As can be seen, the
speed advantage of Rice over Hcompress or GZIP is even greater when
compressing or uncompressing 32-bit images than with 16-bit images. Our
benchmark Linux machine (2.4 GHz AMD Opteron 250 dual core processor), can
Rice-compress a 90 MB 32-bit integer FITS image in about 1 second and can
uncompress the same image in about 1.2 seconds.

\begin{deluxetable}{lcccc}
\tablecaption{Compression statistics for 32-bit integer images
\label{t:32int}
}
\tablewidth{0pt}
\tablehead{
\colhead{ } & \colhead{Rice} & \colhead{Hcompress} & \colhead{Tiled-GZIP} & 
\colhead{Host-GZIP}  
}
\startdata
Compression Ratio      & 3.76 & 3.83 & 2.30 & 2.32 \\
Compression CPU time   & 1.0  & 5.2  & 7.8  & 4.7 \\
Uncompression CPU time & 1.0  & 3.4  & 2.2  & 1.3 \\
\enddata
\end{deluxetable}

\subsection{Special case: Representing Floating-Point Images as Scaled Integers }

\label{s:scaled}
Instead of storing real-valued images using the 32-bit IEEE floating point
number representation, a  widely used FITS convention converts the
floating-point values into scaled integers, where the (approximate)
floating point value is then given by
\begin{equation}
{\sf real\_value} = {\sf BSCALE} \times {\sf integer\_value}  + {\sf BZERO} 
\label{eq:BSCALE}
\end{equation}
and where {\sf BSCALE} and {\sf BZERO} are linear scaling constants given
as keywords in the header of the FITS image.   This is technically a
`lossy' compression technique that  quantizes all the pixel values into a
set of discrete levels, spaced at intervals of 1/{\sf BSCALE}.  Ideally,
the quantization levels should be spaced finely enough so as to not lose
any scientific information in the image,  but without recording each value
with an excessive amount of precision.

Unfortunately, a common practice is to simply scale the image so that the
pixel values span the full 32-bit integer range.  This has the effect of
greatly {\em magnifying} the dispersion in the scaled integer values which
makes them virtually incompressible.

In order to achieve higher compression, a better technique, as described in
detail by \cite{white1999}, is to choose the {\sf BSCALE} value so that the
quantized levels are spaced at some reasonably small fraction of the noise
in the image, such that,
\begin{equation}
{\sf spacing} = 1 / {\sf BSCALE} = \sigma / D 
\label{eq:spacing}
\end{equation}
and where $\sigma$ is calculated from Equation \ref{eq:noise3}.  The number
of noise bits per pixel that are preserved in this case, from  Equation
\ref{eq:nbits}, is simply $\log_2 (D) + 1.792$. In order to achieve the best
compression, data providers should choose the  smallest value of $D$ that
still preserves the required scientific information in the compressed
image.    This can be determined by performing the same  photometric and
astrometric data analysis on both the original and the compressed version
of the image.  The amount of precision that needs to be retained of course
depends on  the particular application (e.g., quick-look or preview images
need much less precision than those intended for full scientific analysis),
but previous experiments \citep[see for example Figure 2 in ][]{white1999}
suggest that values of $D$  in the range of 10 to  100  (about 5 to 8
equivalent noise bits per pixel) may be sufficient to retain all the
scientifically useful information in the image.

\section{Compression of  Floating-point Images}
\label{s:floatimages}

Compressing astronomical images that have 32-bit floating point
format pixel values presents special challenges.  One difficulty 
is that many compression algorithms, including Rice and Hcompress,
can by design only operate on integer data. Another problem is
that in many cases the floating point image is derived from what
was originally a 16-bit integer image (after applying various
calibration operations such as bias subtraction, flat-fielding,
sky subtraction, flux calibration, etc.) and the resulting
expansion from a 16-bit to a 32-bit representation can leave the
less significant bits of the mantissa effectively filled with
uncompressible noise.  One way to mitigate this effect is to
artificially set some of the least significant bits  in every
pixel to zero  which serves to more coarsely quantize  the pixel
values.  This is particularly effective when compressing the
floating point image file with the GZIP utility, because this
reduces the number of different byte patterns present in the
file,  thus increasing the compression efficiency of the 
algorithm.  Some observatories, such as the Swift X-ray and
Gamma-ray satellite, have used this quantization technique in
their processing pipeline to significantly reduce the size  of the
GZIP-compressed files in their data archive.

Rather than directly quantize the floating point values, we use 
an equivalent quantization technique of converting the 
floating-point values into scaled 32-bit integers as described in
section \ref{s:scaled}.  These integers are then  compressed using
the Rice algorithm. The linear scaling parameters are calculated
independently for each tile (row) of the image  so that the
quantization levels are spaced at a user-specified fraction of the
measured noise in the tile. This effectively discards some of the
lower-order bits in the mantissa of the floating point values,
which typically do not contain any significant information. 
Depending on how many bits are discarded, the image compression
ratio of the scaled integer image can be dramatically increased.
As is discussed in section \ref{s:scaled},  however, it is
incumbent upon the user to determine the appropriate scaling level
so as to not degrade the  scientific usefulness of the image. 

\section{Effect of Tiling Pattern on Compression Performance}
\label{s:tilesize}
There are a number of considerations in choosing  an appropriate
tiling pattern when compressing an image. First, the tile must be
sufficiently large for the compression  algorithm to operate
efficiently.  As a general guide line, the lower limit is about
500 pixels for the Rice algorithm and about 2000 pixels for GZIP.
Below these levels, the compression time for the image and the
size of the compressed file both begin to increase.   The
Hcompress algorithm is inherently different from Rice and GZIP in
that the wavelet transform only operates  on 2-dimensional arrays
of data.  At a minimum it requires tiles containing at  least 4
rows of the image, and it reaches near maximum efficiency when the
tiles contain about 16 rows.  For this reason we adopted 16 rows
of the image at a time as the default tiling pattern in our
software when using Hcompress.

The other main consideration when choosing a tile size is how the
software that reads the image will access the pixels. The 2 most
common access methods used by astronomical software  are either to
read the entire array of pixels in the image into computer  memory
all at once, or to read the image sequentially one row at a time. 
In the first case, the specific tiling pattern makes little
difference because the reading routine simply has to uncompress
each and every tile in the image once and pass the array of
uncompressed pixels back to the application program.

If the application program reads the image one row at a time, then
the tiling pattern can have a major effect on the reading speed.
If each tile contains multiple rows of the image (and in the
extreme case, the whole image could be compressed as one big
tile), then  the FITS file reading routine must uncompress the
whole tile in order to extract  a single row.  It would obviously
be very inefficient to repeatedly uncompress the same tile each
time the application program requests the next row of pixels. 
Instead, a recommended implementation strategy is to temporarily
store the most recently accessed uncompressed tile in memory, so
that it is still available in case the application program reads
more pixels from that same tile. This caching  technique adds some
computational overhead, however, so in general the default single
row tiling pattern is more efficient for applications that read an
image row by row. 

A third type of image access occurs in applications that read a
rectangular `cutout' from a much larger compressed image.  In this
case it can be efficient to use a rectangular tile pattern  that
approximates the size of the typical cutout. Only those tiles that
overlap the cutout region will then have to be uncompressed.  This
tiling pattern may be grossly inefficient however, for software
that accesses the image one row at a time, unless a fairly
sophisticated caching mechanism is implemented to store all the
uncompressed tiles along a row.

In summary, the default row by row tiling pattern (or 16 rows at a
time in the case of Hcompress) should work well in most
situations.  The main exception is if the images are very small,
in which case it may be more efficient to compress multiple rows,
or the entire image, as a tile.

\section{Summary}
\label{s:summary}

In this paper we have performed a detailed evaluation of various lossless
compression techniques on a large sample of astronomical images that are
representative of the types of images from many sky survey projects.   Using
optimal compression methods on such images can  make a large difference in
the amount of disk space and network bandwidth  needed to analyze,
archive, and distribute the images.  We focus on lossless compression
techniques because they may be adopted by data providers without any risk
of losing information in the data.  Lossy compression techniques can
provide higher compression and may be appropriate in some situations, but
the data provider must ensure that the required amount of  astrometric and
photometric precision is retained.

As we show in section \ref{s:noisecalc}, the amount by which an image can
be losslessly compressed basically depends on 2 quantities:  the amount of
noise in the image, and on the compression efficiency of the algorithm.
The first quantity can be expressed as the average number of bits of
noise, $N_{bits}$  in each pixel value.   By definition these noise bits
cannot be compressed, therefore the total number of noise bits in the
image sets the lower limit  on the size of the compressed file.  The
efficiency of a compression algorithm is a measure of how well, on
average, it is able to compress  the remaining non-noise bits in each
pixel; we represent this quantity with the letter $K$, in units of
bits per pixel.  The lossless compression ratio
is then simply given by the number of bits per pixel (e.g., 16 or 32)
divided by the sum of $N_{bits}$ and $K$.  

The noise in astronomical images is usually not neatly packed into the
lowest $N_{bits}$ of each pixel value (otherwise they could simply 
be discarded since they contain no useful information). We
measure the {\em equivalent} number of noise bits per pixel from the
fluctuations (i.e.,  noise) in the pixel values in the background regions
of the image.   We use a MAD algorithm that was originally developed to
calculate  the signal-to-noise ratio in spectroscopic data to robustly
estimate the background fluctuations in any image. As we show in the
appendix, if one assumes that the pixel fluctuations have a Gaussian
distribution (which is a reasonable assumption in the CCD images
discussed here) with standard deviation $\sigma$, then the equivalent
number of noise bits per pixel in that image is given  by  ${\log}_2
(\sigma\sqrt{12})$.   

In section \ref{s:synthImages} we constructed synthetic images containing
these 2 different noise distributions (i.e., $N_{bits}$ of noise 
and Gaussian-distributed noise) to verify that the general purpose
compression algorithms used in our study, Rice, Hcompress, and GZIP,
actually behave as expected.   We found that Rice and Hcompress, which
operate on the numerical 16-bit or 32-bit integer value of each pixel, do
indeed show the expected relationship between the amount of noise in the
image and the compression ratio.   In particular, we confirmed that the
offset of $\sqrt{12}$ that was derived in the appendix is needed to bring
the images with the 2 different noise distributions into agreement in
Figures \ref{f:compbit_v_noise16} and \ref{f:compbit_v_noise32}.  The Rice
and Hcompress algorithms also have nearly constant $K$ efficiency
values of 1.2 and 0.8 bits per pixel, respectively, independent of the
amount of noise in the images.   The GZIP algorithm on the other hand
shows quite different behavior, which can be attributed to the fact that it
treats each byte in the 16-bit and 32-bit pixel values as independent
quantities.  The $K$ value for GZIP is much larger (worse) than for
Rice and Hcompress, and it varies depending on the magnitude and the
distribution of the noise in the image.

We then compared the various compression methods on a large  homogeneous
sample of 16-bit integer astronomical images  taken with the NOAO Mosaic
CCD camera, as well as on a smaller diverse sample of images taken with
other instruments.   One of the striking results shown in Figure
\ref{f:rice_v_noise} is how closely the real astronomical images follow
the same tight correlation between compression ratio and noise content as
in the synthetic images.  This demonstrates that the presence of the stars
and faint galaxies in many of these CCD images  has very little effect on
the lossless compression ratio.  This is because these objects  cover only
a small faction of the image area, and the overall compression ratio of
the image is mainly determined by the compressibility of the large
majority of pixels with values close to the local background.  Of course
if the density of the objects in the image is large enough, it will have a
negative impact on the compression ratio of the image, as was seen the
case of a couple images of the central region of a rich globular cluster.

Another interesting result seen in Figure \ref{f:rice_v_noise} is that the
different major types of images (i.e., the bias frames, the short
exposures of calibrations stars, and deeper exposures of the sky and
flat fields) all follow the same continuous relation between compression
ratio and noise content.  They are segregated into different regions in
the figure simply because the average pixel value, and hence the noise, is
distinctly different.  This can have significant consequences when
estimating the storage needs of a data archive,  because the bias frame
and short calibration exposures compress much better than the flat field
and deep exposures of the sky.

The comparison of the compression ratios achieved by the different
algorithms in Figure \ref{f:all_v_noise} shows that Rice and
Hcompress  produce 1.3--1.5 times better compression that GZIP on
16-bit integer astronomical images. Hcompress produces a few percent
better compression than Rice, but it also requires about 3 times more
CPU time.  Rice is 2--3 times faster than the host GZIP utility when
compressing images, and has about the same uncompression speed as
GUNZIP.  Overall Rice provides the best combination of speed and
compression efficiency of the algorithms studied here.   Since the
compression ratios produced by  Rice and Hcompress are already within
75\% to 90\% of an ideal algorithm with $K$ = 0, any
further improvements to the compression algorithms will produce
relatively little gain in the size of the  compressed images.

32-bit integer astronomical images are relatively uncommon, but our
compression tests on one such set of images taken with the NEWFIRM
camera shows that Rice has an even larger performance advantage over
GZIP  in both speed and compression ratios than was the case with the
16-bit images.  One interesting consequence of the fact that the
compression ratio of 32-bit images is twice that of 16-bit images 
(assuming the images contain the same amount of noise), and hence the
compressed files have the same size,  is that there is no disk space
penalty in storing the compressed images as 32-bit integers instead of
16-bit integers.  This might be desired to allow for a greater range
of pixel values during subsequent data processing operations on that
image.

One of the motivations of our work is to publicize and promote  a
better alternative to simply using the GZIP utility to compress FITS
images, as is currently done by most astronomical data providers.  The
FITS tiled image compression format is more efficient for accessing
individual images within a multi-image FITS file, and for reading a
small section from a larger image, and the Rice algorithm produces
much better compression and is faster than GZIP in most cases.  This
new compression format is also supported by major astronomical
software packages such as the CFITSIO library and the ds9 image
display program.  To further encourage the use of this compression
method,  2 open-source image compression programs that were used in
this study (called {\em fpack} and {\em funpack}) are publicly
available for general use from the HEASARC web site
(http://heasarc.gsfc.nasa.gov/fitsio/fpack).  These programs run on
all major computer platforms and  are invoked on the command line,
just like the GZIP and GUNZIP utilities, to compress or uncompress any
FITS image file. Various options  can be specified on the command line
to control the programs.  More information about using these utility
programs is available in the companion user's guide.



\acknowledgments








\appendix

\section{Derivation of the Equivalent Number of Noise Bits}

1) For images with a Gaussian noise distribution (for instance, the readout floor
of a CCD),  we derive the equivalent number of noise bits. Start by assuming N
bits of uniform noise  and average over the range of data numbers ($x$ = DN)
for the expected values of $x$ and $x^2$: 
\begin{eqnarray*}
\langle x \rangle & = & [k (k + 1) /2] / 2^N  \;\;\;\;\;\; (\mbox{$\Sigma$ of DN series, with $k = 2^N - 1$})  \\
  & = & (2^N - 1) / 2  \\
\mbox{and} \;\;\;\;\; \left\langle x^2 \right\rangle & = & [k (k + 1) (2k + 1) / 6] /2^ N  \;\;\;\;\;\; (\mbox{$\Sigma$ of series of squares}) \\
  & = & (2^N - 1) (2^{N+1}- 1) / 6 \\
\mbox{Solve for the variance,} \;\;\;\;\;\;\;\;\;\;\;\;\; \\
\sigma^2 & = & \langle x^2 \rangle - {\langle x \rangle}^2 \\
      & = &   (2^{2N} - 1) / 12 \\
\mbox{In the limit of large $N$:} \;\;\;\;\;\;\;\;\;\; \\
\sigma & = &  2^N / \sqrt{12} \\
\mbox{Solving for $N$ then gives,} \;\;\;\;\;\;\;\;\;\; \\
N_{bits} & = & {\log}_2 (\sigma \sqrt{12}) \\
& = & \log_2 (\sigma) + 1.792 \;\;\;\;\;\;  \mbox{(Equation 4 in section 3)}
\end{eqnarray*}
The factor of $1 / \sqrt{12}$ can be identified as the familiar analog-digital
quantization  noise \citep{janesick2001}.  The same result also follows with
continuous variables by integrating the second moment of a stepwise probability
density symmetrically over $2^N$ quanta. This discrete derivation makes the
non-linear limiting behavior at small values of N evident.

2) Our quantity $N_{bits}$ is equivalent to the Shannon entropy  (really
just a definition of terms). Entropy is a sum over all possible states of some
discrete random  variable (pixels in our case), and depends only on the
probabilities of each state, not on  their values:
\begin{eqnarray*}
H & = & -\sum p_i \log p_i
\end{eqnarray*}
For uniform noise, all states are equally probable, so $p_i = 1 / 2^N$:
\begin{eqnarray*}
H & = & -\sum 2^{-N} \log_2 2^{-N} \\
& = & + N \left(\sum 2^{-N}\right) 
\end{eqnarray*}
The sum equals unity since it is over $2^N$ states where $N = N_{bits}$,  
thus $H = N_{bits}$.

3) For a continuous random variable with probability $f(x)$, the differential
entropy is:
\begin{eqnarray*}
h & = & - \int f(x) \log f(x) dx
\end{eqnarray*}
There are caveats as with any integral, e.g., one must consider whether a
solution exists in each case. More fundamentally, the differential case (as the
name suggests) provides a measure of entropy that can be understood only
relative to a particular coordinate frame  -- unlike the discrete entropy that
provides an absolute measure of randomness.

The differential entropy of a Gaussian probability density function is
\citep{shannon1948, cover1991, roger1994}:
\begin{eqnarray*}
h_g & = & \log_2 \sqrt{2 \pi e \sigma^2} \\
& = & \log_2 \sigma + 2.047
\end{eqnarray*}

4) How does our equation 4 correspond to the previous result? To answer this,
consider Massey's bound on the discrete entropy \citep[as derived in]
[referencing unpublished work by J. Massey and independently by F. Willems]{cover1991}.
Starting with the fact that the Gaussian probability distribution has the maximum entropy
for a given variance, the differential expression leads to a bound on the entropy of a
discrete random variable (a similar inequality holds for all permutations):
\begin{eqnarray*}
H & \leq & (1/2 ) * \log \left[ 2 \pi e \left(\sum p_ii^2 - \left( \sum p_ii\right)^2 + 1/12\right) \right]
\end{eqnarray*}
As before, assume N bits of uniform noise such that $p_i = 2^{-N}$. Use the same expressions
to sum the two series as above, but with limits from 1 to $2^N$ (i.e., k = $2^N$ since we are
counting states here, not calculating statistical moments as in equation 4):
\begin{eqnarray*}
H & \leq & (1/2 ) * \log \left[ 2 \pi e \left(p_i \sum i^2 - {p_i}^2 \left( \sum i \right)^2 + 1/12 \right) \right] \\
& = & (1/2 ) * \log [ 2 \pi e (p_i (k (k+1)(2k+1)/6) - {p_i}^2 (k(k+1)/2)^2 + 1/12) ] \;\;\; (etc.) \\
& = & (1/2 ) * \log [ 2 \pi e (2^{2N})/12] \\ 
& = & N + \log_2 \sqrt{\pi e / 6} = N + 0.2546
\end{eqnarray*}
However, we have already shown the identity $H = N_{bits}$, such that:
\begin{eqnarray*}
H & = & N_{bits} \leq  N_{bits} + 0.2546
\end{eqnarray*}
The units of bits for the entropy are fixed when choosing the base-2 logarithm. It is
beyond the scope of the current work to ponder the origin of this one-quarter bit offset. It
is sufficient to note that the inequality is satisfied, and further to attribute the difference
of $\log_2 \sqrt{\pi e / 6}$ to a coordinate transformation in the differential entropy.

5) In particular, to motivate equation 4 we can remove this same coordinate offset from
the Gaussian differential entropy ($h_g$) to obtain a measure of the Shannon entropy ($H_g$):
\begin{eqnarray*}
h_g & = & \log_2 \sqrt{2 \pi e \sigma^2} \\
H_g & = & \log_2 \sqrt{2 \pi e \sigma^2} - \log_2 \sqrt{ \pi e / 6} \\
    & = & \log_2 \sqrt{2 \pi e \sigma^2 / (\pi e / 6)} \\
\mbox{Therefore,} \;\;\;\;\;\; H_g & = & \log_2 ( \sigma \sqrt{12}) = N_{bits}
\end{eqnarray*}
In summary, equation 4 was derived by calculating the variance of $N_{bits}$ of
uniform random noise. A coordinate transformation of the differential entropy 
of a Gaussian leads to the same equation. (A Gaussian maximizes the entropy for
a given variance, suggesting this equivalence is an identity.) Given $N_{bits}$,
one can estimate  the variance. On the other hand, given the statistical
variance of some random variable (easily estimated from the pixel values of an
astronomical image), an estimate follows of the equivalent number of effective
noise bits, that is, of the image's entropy in bits. With an estimate of entropy
in hand, the compression factor is calculable.

6) Figures 1 and 2 demonstrate empirical agreement of synthetic Gaussian data
with the $N_{bits}$  relation for both 16-bit and 32-bit integer pixels. Figures
3 and 7 empirically confirm this relation  for the Gaussian read-noise dominated
background of real world optical and infrared data sets, and further, for both
on-sky and calibration data products.





\end{document}